\documentclass  [12pt] {article}

\def\vp{\vec{p}\, {}^2}
\newcommand{\dd}[2]{\frac{\partial #1}{\partial #2}}
\begin{document}
\title{%
Velocity of particles in Doubly Special Relativity}
\author{M.\ Daszkiewicz\thanks{e-mail address marcin@ift.uni.wroc.pl},~~K.\
Imilkowska\thanks{e-mail address kaim@ift.uni.wroc.pl},~~and J.\
Kowalski--Glikman\thanks{e-mail address
jurekk@ift.uni.wroc.pl}~~\thanks{Partially supported by the   KBN
grant 5PO3B05620}\\  \\ {\em Institute for Theoretical
Physics}\\ {\em University of Wroc\l{}aw}\\ {\em Pl.\ Maxa Borna 9}\\
{\em Pl--50-204 Wroc\l{}aw, Poland}} \maketitle
\begin{abstract}
Doubly Special Relativity (DSR) is a class of theories of
relativistic motion with two observer-independent scales. We
investigate the velocity of particles in DSR, defining velocity as
the Poisson bracket of position with the appropriate hamiltonian,
taking care of the non-trivial structure of the DSR phase space.
We find the general expression for four-velocity, and  we show
further that the three-velocity of massless particles equals 1 for
all DSR theories. The relation between the boost parameter and
velocity is also clarified.

\end{abstract}
\clearpage

\section{Introduction}

There is a rapidly growing interest in a research programme, which
might be called ``Quantum Special Relativity'', i.e. a class of
theories of kinematics that differ in their predictions from that
of Special Relativity in the regime of ultra-high energies. These
possible differences might be understood as traces of still
unknown Quantum Gravity theory present even in the regime of
negligible gravitational field. It is feasible that predictions of
such generalizations of Special Relativity, like dependence of the
speed of massless particles on momentum they carry, might be
experimentally tested in the near future experiments.

Two alternative classes of ``Quantum Special Relativity'' have
been recently attracting  attention. In one, one argues that
Lorentz invariance is broken at high energies due to string
\cite{Colladay:1998fq}, loop-quantum gravity \cite{Gambini:1998it}
effects, or due to existence of the cosmological preferred frame
\cite{Jacobson:2001yj}. In the second, proposed by Amelino-Camelia
\cite{Amelino-Camelia:2000ge}, \cite{Amelino-Camelia:2000mn},
called Doubly Special Relativity (DSR), one assumes that the ten
dimensional algebra of physical symmetries (rotations, boosts, and
translations) is still present, but is deformed in a way so as to
possess two observer-independent scales. In this paper we will
concentrate on this second possibility only.

The construction of the first specific model of DSR, called DSR1, presented in
\cite{Kowalski-Glikman:2001gp} and \cite{Bruno:2001mw} borrowed a lot from the
earlier investigations in quantum Hopf deformations of Poincar\'e algebra, the
so-called $\kappa$-Poincar\'e algebra (see e.g., \cite{lunoruto}, \cite{maru}).
It turned out however that there exist another DSR models: the first one,
called nowadays DSR2 was formulated by Magueijo and Smolin
\cite{Magueijo:2001cr}, and it was soon realized that there exist a whole class
of DSR theories \cite{Kowalski-Glikman:2002we}, \cite{Kowalski-Glikman:2002jr}.

In the majority  of the (especially early) literature devoted to
DSR, Doubly Special Relativity is defined only as a theory based
on energy-momentum sector. This is {\em not} what we understand by
DSR in this paper. In our view, a DSR theory is defined by a set
of commutators describing the whole of the phase space of the
system. There are two systematic and equivalent ways of deriving
such a set in any particular model\footnote{Some other methods of
deriving consistent phase space may exist, of course. We are not
aware however of any nontrivial alternative to the procedure we
make use of.}. The first is based on the Hopf algebra structure
that can be built on the energy--momentum algebra, and the use of
the so called Heisenberg double construction \cite{luno},
\cite{crossalg},
 \cite{Kowalski-Glikman:2002jr}. Equivalently, one can get the same
phase space by making use of the geometric, de Sitter picture of DSR,
\cite{Kowalski-Glikman:2002ft}, \cite{jsnew}.
\newline

Doubly Special Relativity is a theory of particle kinematics, and the proper
understanding of the concept of velocity in such a theory is of course an
important step towards full understanding of it. The first attempt to analyze
the notion of velocity has been made already in the early days of the
$\kappa$-Poincar\'e theory in \cite{Lukierski:1993wx}, from the DSR perspective
this problem has been investigated, among others, in
\cite{Kowalski-Glikman:2001px}, \cite{Lukierski:2002fd},
\cite{Amelino-Camelia:2002tc}, \cite{Kosinski:2002gu}, \cite{mignemi2003}.

The starting point of our investigations reported here consists of
two major assumptions: that velocity is defined as the Poisson
bracket of position with deformed relativistic hamiltonian (see
also \cite{Lukierski:1993wx}), and that to compute this bracket
one must take into account the nontrivial phase space structure of
DSR theories.  We show that in all DSR theories four velocities
transform as  standard Lorentz vectors, and that the three
velocities of massless particles equal one. This general statement
will be justified in Section 3; before turning to that, we present
our method on the specific example of DSR1.

\section{Particle velocity in DSR1}

Before starting our investigations let us state clearly what our
assumptions are. We define four velocity as the Poisson bracket,
based on a DSR phase space structure, of positions with an
appropriate hamiltonian. Let us explain why we decided to use
Poisson brackets instead of commutators, usually employed in the
DSR literature. The reason is quite simple: the use of commutators
implies that the relevant objects are operators acting on some
Hilbert space. Since no investigations of such operators'
properties  (functional analysis) has been performed so far,
working with the commutator algebra is equivalent to phrasing
results in terms of Poisson brackets. Since we will make use of
the expression for three-velocity as a function of four velocities
$v^i = {\dot x}^i/{\dot x}^0$ whose meaning for $x^0$ being an
operator is not clear, it is just safer to work with Poisson
brackets. For this reason we will confirm our discussion to
classical particles and not quantum matter waves. In our
investigations in this section we will also postulate a particular
form of relativistic hamiltonian. We will argue that such form is
natural in the next section devoted to general properties shared
by all DSR theories.

In all the DSR theories (contrary to the statements that can be sometimes found
in the literature), the Lorentz algebra of rotations $M_i$ and boosts $N_i$ is
exactly the same as in Special Relativity\footnote{Let us stress again that
$[\ast,\ast]$ denotes the Poisson bracket and {\em not} the commutator.}
$$
[M_i, M_j] =  \epsilon_{ijk} M_k, \quad [M_i, N_j] =  \epsilon_{ijk} N_k,
$$
\begin{equation}\label{1}
  [N_i, N_j] = - \epsilon_{ijk} M_k,
\end{equation}

In the DSR1 theory the momenta transform under action of boosts as follows
\begin{equation}\label{2}
   \left[N_{i}, p_{j}\right] =   \delta_{ij}
 \left( {\kappa\over 2} \left(
 1 -e^{-2{p_{0}/ \kappa}}
\right) + {1\over 2\kappa} \vec{p}\,{}^{ 2}\, \right) -  {1\over \kappa}
p_{i}p_{j} ,
\end{equation}
and
\begin{equation}\label{3}
  \left[N_{i},p_{0}\right] =  p_{i}.
\end{equation}
The first Casimir of this theory equals
\begin{equation}\label{4}
{\cal C}  = \left(2\kappa \sinh \left(\frac{p_0}{2\kappa}\right)\right)^2 -
\vec{p}\,{}^2\, e^{p_0/\kappa}=m^2.
\end{equation}
Let us now turn to description of the phase space of DSR1. As for all DSR
theories we have
\begin{equation}\label{5}
  [x_0, x_i] = -\frac{1}{\kappa}\, x_i, \quad [x_i, x_j] = 0.
\end{equation}
As shown in \cite{jsnew} there are infinitely many phase spaces compatible with
the DSR1 boost transformations (\ref{3}), (\ref{4}) and the brackets (\ref{5}).
Here we will describe only the two most simple ones for whose the cross
brackets take the following form
$$ [p_0,
x_0] = -1, \quad [p_i, x_0] =  \frac{1}\kappa\, p_i,
$$
\begin{equation}\label{6}
[p_i, x_j] =  \delta_{ij} \, e^{- 2p_0/\kappa}
-\frac{1}{\kappa^2}\left(\vec{p}\,{}^{ 2}\, \delta_{ij} - 2 p_{i}p_{j}\right),
\quad [p_0, x_i] = -\frac{2}\kappa\, p_i
\end{equation}
or, following \cite{maru}, \cite{Lukierski:1993wx}
\begin{equation}\label{6a}
  [p_0, x_0] = 1, \quad [p_i, x_0] =  -\frac{1}\kappa\, p_i, \quad [p_i, x_j] =-
 \delta_{ij} , \quad [p_0, x_i] =0.
\end{equation}
We choose the hamiltonian to be
\begin{equation}\label{7} { \cal H} =
\kappa^2\, \cosh \frac{p_0}\kappa  - \frac{\vec{p}\,{}^2}{2} \, e^{
\frac{p_0}\kappa}.
\end{equation}
This hamiltonian has the large $\kappa$ limit
$$
{ \cal H} \sim \kappa^2 +\frac12\left(p^2_0 -\vec{p}\,{}^2\right) + \ldots
$$
i.e., it reduces in this limit to the standard hamiltonian of relativistic
particle (up to the irrelevant constant shift). Let us now define the four
velocities in the standard way as the bracket
\begin{equation}\label{8}
 u_0 \equiv  \dot x_0 =  [x_0,{\cal H}], \quad u_i \equiv  \dot x_i =  [x_i,{\cal H}]
\end{equation}
In calculating the brackets in (\ref{8}) one should carefully take care of the
nontrivial phase space structure of DSR1 (\ref{6}) or (\ref{6a}) (in accordance
with the scheme presented in \cite{Lukierski:1993wx})
$$
  [x_0,{\cal H}] \equiv \dd{{\cal H}}{p_0}\, [x_0, p_0] + \dd{{\cal H}}{p_i}\, [x_0,
  p_i],
  $$
  \begin{equation}\label{9} [x_k,{\cal H}] \equiv \dd{{\cal H}}{p_0}\, [x_k, p_0] + \dd{{\cal H}}{p_i}\, [x_k,
  p_i].
\end{equation}
The second set of Hamilton equations is quite simple
\begin{equation}\label{10}
\dot p_\mu =0
\end{equation}
because in all DSR theories the hamiltonian depends on momenta only, and
momenta have vanishing bracket among themselves. Notice that this property
guarantees that in DSR free motion is uniform. One should note at this point
that for this reason it seems that the so called ``twisted phase spaces'',
investigated recently in \cite{czerhoniak2003}, which would lead to non-uniform
motion of free particles, are likely not to be physical.

It is easy to derive the expression for four velocity, following our general
prescription (\ref{9}). We find
\begin{eqnarray}
{u_0} &=& \kappa\, \sinh \frac{p_0}\kappa + \frac{\vec{p}\,{}^2}{2\kappa}\,
e^{  \frac{p_0}\kappa} \nonumber\\
u_i &=&   p_i \, e^{  \frac{p_0}\kappa}    \label{11}
\end{eqnarray}
 in the  case of phase space (\ref{6}) and
\begin{eqnarray}
{u_0} &=& -\kappa\, \sinh \frac{p_0}\kappa -\frac{\vec{p}\,{}^2}{2\kappa}\,
e^{  \frac{p_0}\kappa} \nonumber\\
u_i &=&   - p_i \, e^{  \frac{p_0}\kappa}    \label{12}
\end{eqnarray}
for the phase space (\ref{6a})\footnote{In this particular case
our result agrees with the one
 reported in \cite{Lukierski:2002fd}, \cite{Kosinski:2002gu}.}. We see that the four velocities differ by the
overall sign, but that in both cases the expression for three-velocity is
exactly the same and reads
\begin{equation}\label{13}
 v_i = \frac{u_i}{u_0} = p_i\,
 \left(\frac\kappa2\, \left(1 -e^{-2  {p_0}/\kappa}\right) +\frac{\vec{p}\,{}^2}{2\kappa}\right)^{-1}
\end{equation}
Using the expansion of the mass-shell condition for massless particles
(cf.~eq.~(\ref{4}))
\begin{equation}\label{19}
0 =\kappa^2\left(1 - e^{-p_0/\kappa} \right)^2 - \vec{p}\,{}^2
\end{equation}
we find
\begin{equation}\label{20}
 v^{(m=0)}_i =   \frac{p_i}\kappa\,
 \left( 1 -e^{-  {p_0}/\kappa}\right)^{-1}.
\end{equation}
Using (\ref{19}) again we find that the speed of massless particles
\begin{equation}\label{20a}
 v^{(m=0)} = \left|v^{(m=0)}_i\right| = 1
\end{equation}
as in Special Relativity. Of course, there are deviations from Special
Relativistic results in the case of massive particles. Indeed, in the massive
case  we have with the help of (\ref{4})
\begin{equation}\label{20b}
 v_i^{(m)} = p_i\, \left[\kappa\left(1-e^{-p_0/\kappa}\right) - \frac{m^2e^{-p_0/\kappa}}{2\kappa}\right]^{-1}
\end{equation}
Note that the speed-of-light of a massive particle, i.e., the maximal speed of
massive particle carrying infinite energy equals again
\begin{equation}\label{20c}
  v^{(\infty)} =1
\end{equation}
because $p_0=\infty$ corresponds to $|\vec{p}|=\kappa$, as it follows simply
from (\ref{4}).

It is worth noticing that if one derives the transformation rules for four
velocities (\ref{11}), (\ref{12}) under action of boosts, using expressions
(\ref{2}), (\ref{3}), one finds that they transform as standard Lorentz four
vectors of Special Relativity. This is not an accidental  property of DSR1, in
fact it holds for all DSR theories. Let us turn therefore to general
formulation and properties of such theories.

\section{Velocity in DSR theories -- generalities}

The results derived in the previous section turn out to be valid not only for
DSR1, but in fact for all DSR theories. In order to see that let us recall the
geometric, de Sitter space formulation of the DSR theories presented in
\cite{Kowalski-Glikman:2002ft}. The starting point here is a five dimensional
manifold of Minkowski signature
\begin{equation}\label{21}
ds^2 = g^{AB}d\eta_A d\eta_B = -d\eta_0^2 + d\eta_i^2 + d\eta_4^2.
\end{equation}
In this space the four dimensional de Sitter space is imbedded by
\begin{equation}\label{22}
 -\eta_0^2 + \eta_1^2+ \eta_2^2+ \eta_3^2+ \eta_4^2 =\kappa^2,
\end{equation}
We split the ten dimensional algebra of isometries of de Sitter space
(\ref{22}) into the six dimensional Lorentz algebra (\ref{1}) and the
remainder, which we identify with positions $x_\mu$ satisfying (\ref{5}). The
remaining brackets are
\begin{equation}\label{23}
 [M_i, \eta_j] = \epsilon_{ijk}\, \eta_k, \quad [M_i,
 \eta_0]=0,\quad
[M_i, \eta_4] =0
\end{equation}
\begin{equation}\label{24}
 [N_i,\eta_j] =  \delta_{ij}\, \eta_0, \quad [N_i, \eta_0] =  \eta_i,\quad [N_i, \eta_4] =0
\end{equation}

\begin{equation}\label{25}
  [x_0,\eta_4] = \frac{1}\kappa\, \eta_0, \quad [x_0,\eta_0]
= \frac{1}\kappa\, \eta_4, \quad [x_0,\eta_i] = 0,
\end{equation}
\begin{equation}\label{26}
  [x_i, \eta_4] =\frac{1}\kappa\, \eta_i, \quad [x_i, \eta_0] =\frac{1}\kappa\, \eta_i, \quad
  [x_i, \eta_j] = \frac{1}\kappa\,
\delta_{ij}(\eta_0 - \eta_4),
\end{equation}

The main result of \cite{Kowalski-Glikman:2002ft}, based on the previous
investigations reported in \cite{Kowalski-Glikman:2002we} and
\cite{Kowalski-Glikman:2002jr} is that any DSR theory can be represented as a
particular coordinate system on the de Sitter space (\ref{22}), i.e., the
mapping from the space of physical momenta $p_\mu$ to $\eta_A$ satisfying
(\ref{22}). For example, in the case of the DSR1 theories with phase spaces
(\ref{6}), (\ref{6a}) this mapping takes the form
\begin{eqnarray}
{\eta_0} &=& \pm \left(\kappa\, \sinh \frac{p_0}\kappa +
\frac{\vec{p}\,{}^2}{2\kappa}\,
e^{  \frac{p_0}\kappa}\right) \nonumber\\
\eta_i &=&  \pm \left( p_i \, e^{  \frac{p_0}\kappa} \right)\nonumber\\
{\eta_4} &=&  \kappa\, \cosh \frac{p_0}\kappa  - \frac{\vec{p}\,{}^2}{2\kappa}
\, e^{  \frac{p_0}\kappa}.   \label{27}
\end{eqnarray}
Let us observe now that in view of eqs.~(\ref{22}) and (\ref{24}) $\kappa \,
\eta_4$ is the most natural candidate for relativistic hamiltonian. Indeed it
is by construction Lorentz-invariant, and reduces to the standard relativistic
particle hamiltonian in the large $\kappa$ limit. Indeed, using the fact that
for $p_\mu$ small compared to $\kappa$, in any DSR theory $\eta_\mu \sim p_\mu
+ O(1/\kappa)$ we have
\begin{equation}\label{a}
 \kappa\eta_4 = \kappa^2\sqrt{1 + \frac{p_0^2 -\vp}{\kappa^2}} \sim \kappa^2 + \frac12\left(p_0^2 -\vp\right) +
 O\left(\frac1{\kappa^2}\right)
\end{equation}

Then it follows from eqs.~(\ref{25}), (\ref{26}) that $\eta_\mu = [x_\mu,
\kappa\eta_4]$ can be identified with four velocities $u_\mu$. The Lorentz
transformations of four velocities are then given by eq.~(\ref{24}) and are
with those of  Special Relativity. Moreover, since
\begin{equation}\label{28}
 u_0^2 - \vec{u}\,{}^2 \equiv {\cal C} = m^2
\end{equation}
by the standard argument the three velocity equals $v_i = u_i/u_0$ and the
speed of massless particle equals $1$.  Let us stress here once again that this
result is DSR model independent, though, of course, the relation between three
velocity of massive particles and energy they carry depends on a particular DSR
model one uses.

Thus if in the time-of-flight experiment (see e.g.,
\cite{Amelino-Camelia:2002vw} for recent review), will measure  time gap
between arrivals of photons emitted from a distant source, and carrying
different energies,  this would falsify the construction of the DSR models
presented above.

Note finally that as the result of eq.~(\ref{28}) and the definition of three
velocity as $v_i = u_i/u_0$, the rule of adding the latter in any DSR theory is
identical with the standard rule of Special Relativity. It is however an open
question if this can be consistently extended to the rule of addition of
momenta, following the considerations of Lukierski and Nowicki
\cite{Lukierski:2002df} and  Judes and Visser \cite{Judes:2002bw}.

\section{Comments and concluding remarks}

The main result of our investigations reported here is that if the phase space
of DSR theories is constructed in the way suggested by geometric picture of de
Sitter space, the speed of massless particle equals 1. There is a number of
simple observations one can make. First, in the scheme adopted here, for any
DSR theory the velocity is just
\begin{equation}\label{28a}
v_i = \frac{u_i}{u_0} = \frac{\eta_i(p)}{\eta_0(p)}
\end{equation}
which provides the velocity--momentum relation for an arbitrary DSR theory,
since in any DSR theory the variables $\eta_\mu$ are functions of momenta,
given by its definition. It follows also that the boost parameter $\xi$ is
related to velocity in exactly the same way as in the Special Relativity
\begin{equation}\label{29}
 \tanh\xi = v
\end{equation}
This can be easily seen by realizing that the brackets (\ref{24}) are
equivalent to the equations (for boost acting in $3$rd direction, say)
\begin{equation}\label{30}
 \frac{d\eta_3}{d\xi} = \eta_0, \quad \frac{d\eta_0}{d\xi} = \eta_3,
\end{equation}
from which (\ref{29}) immediately follows. To obtain the corresponding equation
for  dependence of momenta $p_\mu$ on rapidity in a particular DSR theory,
defined by a particular functions $\eta_A(p_\mu)$, one uses eq.~(\ref{30}) and
the Leibnitz rule. But this does not change the relation (\ref{29}) where the
right hand side can be taken  a given function of momenta.
\newline

It is interesting to compare our result with the calculation of the group
velocity of wave packets presented in \cite{Amelino-Camelia:2002tc}. In this
paper the authors perform their computations using essentially only the
noncommutative structure of $\kappa$-Minkowski space-time (\ref{5}), the
particular form of non-commutative differential calculus, and a natural
ordering of plane waves. Their result, that the group velocity $v^{(g)} =
\partial p_0/\partial p$ (where the derivative is taken on the mass-shell, i.e.\ assuming that
eq.~(\ref{4}) holds,) should be therefore  valid for all DSR theories. The
clear disagreement of their result with
 the one presented in this paper deserves further studies, since it seems to
 indicates that in the framework of DSR the
 behavior of matter waves may differ from that of particles. Specifically, this discrepancy means that either
 naively constructed wave
 packets could not represent point particles, so that the latter are represented by a non-linear combination of plane waves,
 or that one of these two notions (linear wave packet and/or point particles) just does not make physically sense in the
 DSR theories.

 To conclude let us stress that the final judgement on possible dependence of velocity on
 energy will be made by near future experiments. We would like to note however that if it turns out that
 our result is correct, i.e., if in the DSR theories we still have to do (as in Special Relativity) with
 universal speed of physical signals carried by massless particles, this will constitute an important
 information, making it easier to formulate Doubly Special Relativity operationally.

\section*{Acknowledgement}
We would like to thank J.~Lukierski for his valuable comments on
the draft of this papers.

\end{document}